\documentclass[lettersize,journal]{IEEEtran}
\usepackage{amsmath,amssymb,amsfonts}
\usepackage{algorithmic}
\usepackage[ruled]{algorithm2e}
\usepackage{array}
\usepackage{textcomp}
\usepackage{stfloats}
\usepackage{url}
\usepackage{color}
\usepackage{verbatim}
\usepackage{graphicx}
\usepackage{multirow}
\usepackage{array}
\usepackage{booktabs}
\usepackage{graphicx}
\usepackage{parskip}

\usepackage[numbers]{natbib}
\usepackage{bm}
\usepackage{mathtools}

\usepackage{subfigure}
\usepackage{makecell}
\usepackage{enumitem}

\usepackage{bibspacing}
\graphicspath{ {./fig/} }

\usepackage{flushend}

\usepackage{stackengine}

\def\BibTeX{{\rm B\kern-.05em{\sc i\kern-.025em b}\kern-.08em
    T\kern-.1667em\lower.7ex\hbox{E}\kern-.125emX}}
\usepackage{balance}
\begin{document}

\title{Trust-as-a-Service: Intelligent Collaboration Orchestration via Model Context \\ Protocol-Aided Agentic AI} 
\author{
Botao~Zhu,~\IEEEmembership{Member,~IEEE} and Xianbin~Wang,~\IEEEmembership{Fellow,~IEEE}
\thanks{

B. Zhu and X. Wang (Corresponding author) are with the Department of Electrical and Computer Engineering, Western University, Canada. 
}
}

\maketitle

\begin{abstract}

As future networked systems increasingly rely on collaborative task execution among distributed devices, trust becomes essential for identifying reliable collaborators whose capabilities and resources match task-specific needs. However, diverse task needs, limited task-owner knowledge, and complex inter-device relationships make it challenging to evaluate the trustworthiness of potential collaborators and to select suitable collaborators for task completion. To address these challenges, this paper proposes Trust-as-a-Service (TaaS), an intelligent collaboration orchestration paradigm that enables trust evaluation and collaborator selection to be autonomously tailored to different task needs. To realize TaaS, we develop a Model Context Protocol (MCP)-aided agentic AI framework. The central server-side agent autonomously performs trust-related operations according to task-specific needs and delivers trust assessment services to task owners through a unified interface. Meanwhile, device-side agents expose their capabilities and resources via MCP servers, allowing devices to be dynamically discovered, evaluated, engaged, and released to form task-specific collaborative units. Experimental results demonstrate that the proposed TaaS achieves 100\% collaborator selection accuracy, along with high reliability and resource-efficient task completion.
\end{abstract}

\section{Introduction}

\IEEEPARstart{T}{he} rapid evolution of wireless technologies toward 6G, coupled with their unprecedented convergence with collaborative computing and artificial intelligence (AI), is enabling a wide range of networked Internet of Things (IoT) systems and applications. Task forms are becoming increasingly diverse and dynamic--from immersive interaction and digital twins to collaborative robotics, autonomous vehicle fleets, and cross-domain distributed sensing with real-time intelligent processing~\cite{11024060}. Consequently, task execution is no longer a localized behavior of individual devices, but a system-level cooperative process that spans distributed devices and heterogeneous resources toward common task goals~\cite{10508191}. In such cooperative processes, a critical prerequisite is the ability to identify and select reliable collaborators whose capabilities genuinely match the task demands. However, due to task complexity, system dynamics, 
and device heterogeneity, traditional security- and privacy-based collaboration paradigms--which focus primarily on authentication and data protection--are insufficient to assess whether a collaborator can 
effectively fulfill specific task requirements.

Within this context, trust emerges as a new mechanism for enabling effective collaboration among distributed devices to achieve reliable task completion in complex IoT systems. Beyond traditional security and privacy guarantees, trust provides a holistic tool for evaluating potential collaborators' reliability for specific task goals by analyzing their semantic trust information, such as device states, historical behaviors, communication and computation resource availability, interaction feedback, and other context-dependent factors~\cite{11395598}. To be effective, trust should serve as a built-in system mechanism that supports the full task collaboration lifecycle--spanning task goal comprehension, semantic trust information collection, trust evaluation, collaborator selection, task execution, and result feedback~\cite{9866814}. However, the dynamic and heterogeneous nature of these stages demands a level of adaptability and contextual reasoning that traditional machine learning and deep learning methods struggle to provide~\cite{9834331}. In contrast, large language model (LLM)-driven agentic AI offers a promising pathway for realizing such intelligent trust mechanisms. Equipped with capabilities in goal-directed reasoning, contextual understanding, multi-stage planning, and tool invocation~\cite{10558819}, agentic AI can dynamically organize information, coordinate resources, and adapt strategies in accordance with specific task requirements~\cite{10638533}. By deploying agentic AI on devices within a collaborative IoT system, devices can expose local capabilities, participate in trust-related interactions, and coordinate with one another for task execution. However, given the high complexity of collaborative systems, the autonomous operation of individual devices, and the functional independence across network layers, implementing an agentic AI-driven intelligent trust mechanism to support task-specific collaboration and ensure effective task completion must address the following challenges.

\textit{How to leverage distributed and diverse semantic trust information with varying quality for reliable trust evaluation?} Trust evaluation relies on diverse semantic trust information from distributed devices, including historical behaviors, capability states, and interaction records, which makes it difficult for a single device to access all of this information~\cite{10103199}. In addition, such information is generated under diverse contexts and varies in quality, necessitating situation-aware analysis and calibration by a credible third party. Furthermore, devices may be reluctant to directly expose their information to other peers due to privacy and competitive concerns. However, they are more willing to share information with a neutral and credible intermediary that is not a direct competitor, making a centralized trusted layer not only necessary but also feasible for collecting and managing such information in a unified way. Finally, semantic trust information across devices is inherently interdependent, necessitating a global perspective to properly analyze and interpret it for accurate trust evaluation. The distributed nature of semantic trust information, its heterogeneous quality, independent ownership, and inter-device dependencies collectively indicate that trust evaluation should not be undertaken by individual devices independently. Instead, trust evaluation should be implemented as a centralized service to more effectively leverage distributed semantic trust information.

\textit{How to implement a need-driven trust evaluation process as a service cost-effectively?} Different tasks differ fundamentally in their collaborator requirements. For example, a latency-sensitive task may prioritize a collaborator with fast processing speed and real-time responsiveness, while a data-intensive task may require a collaborator with long-term availability and a reliable data integrity track record~\cite{10525057}. Applying a uniform, exhaustive trust evaluation strategy across all tasks would inevitably activate irrelevant evaluation dimensions, incurring unnecessary computational and communication overhead without delivering any corresponding benefit~\cite{8897627}. A more cost-effective approach is to provide trust evaluation as a service to task owners, taking task needs as input and dynamically selecting only the necessary evaluation dimensions and decision strategies. This need-driven provisioning ensures that evaluation effort is directed to task-relevant dimensions, thereby reducing redundant overhead. Furthermore, by aligning task-specific needs with system capabilities, trust evaluation as a service can balance evaluation effectiveness and service cost, thereby enabling concurrent trust evaluation across diverse tasks in complex systems.

\textit{How to enable autonomous collaboration among distributed devices for task-specific execution?} In existing networks, devices are typically organized into collaborative groups based on static criteria such as resources, geographic proximity, protocol compatibility, or fixed roles~\cite{10989563}. However, as different tasks impose distinct capability requirements, static composition cannot adapt to such variability, leading to underutilized resources, rigid structures, and poor cross-device coordination. Therefore, a new composition model is required, in which devices are autonomously orchestrated into temporary collaborative units tailored to each task and dissolved upon completion. In agentic AI-driven systems, this becomes feasible--each device, functioning as an autonomous agent, is capable of perceiving its environment, reasoning about tasks, and acting independently, providing the intelligence needed to participate in such flexible collaboration. However, realizing this potential requires that devices be discoverable in real time, evaluable on demand, and readily engageable in task execution. A straightforward approach is for each device to maintain peer records (e.g., resources and capabilities) for on-demand collaboration, but this incurs prohibitive maintenance overhead as the system scales. Instead, devices can expose their own resources, capabilities, and status through standardized interfaces, making them discoverable, queryable, and callable across the system. This standardized exposure provides the foundation for trust-guided collaborator selection, collaborator engagement, and runtime coordination, allowing the system to efficiently form, coordinate, and dissolve collaborative units on demand.

To address the aforementioned challenges, we propose Trust-as-a-Service (TaaS). The key idea is to abstract complex trust mechanisms into a unified, system-wide service, enabling efficient utilization of distributed semantic trust information, need-driven provisioning of trust evaluation services, and trust-guided collaborator organization. To implement TaaS, we design an agentic AI-driven framework built upon the Model Context Protocol (MCP). At its core, a server-side intelligent agent provides the trust evaluation service to all task owners through a standardized interface and autonomously carries out trust-related operations tailored to task-specific needs. On the collaborator side, device agents publish their capabilities and semantic trust information through MCP servers, enabling them to be dynamically discovered, evaluated, engaged, and released as task demands dictate. As a result, the proposed framework ensures task-specific, consistent, efficient, and trustworthy collaboration. 
The main contributions of this paper are summarized as follows.
\begin{itemize}
    \item We propose a novel TaaS paradigm that transforms trust evaluation into a system-level service, providing task owners with global insights for high-quality trust assessment and efficient utilization of distributed semantic trust information.

     \item We design an MCP-aided agentic AI framework to realize TaaS, where the server-side agent provisions trust evaluation services through standardized MCP interfaces and device-side agents expose resources, capabilities, and tools for on-demand interaction.

    \item  We introduce a need-driven trust assessment strategy that interprets task needs and derives task-specific trust evaluation dimensions, enabling task-tailored trust assessment across heterogeneous tasks.

   \item  We enable trust-guided organization of distributed collaborators via MCP servers, supporting collaborator selection, task assignment, execution feedback, and the on-demand formation and dissolution of task-specific collaborative units.
    
\end{itemize}

\begin{figure*}[!]
\centering
\includegraphics[scale=1]{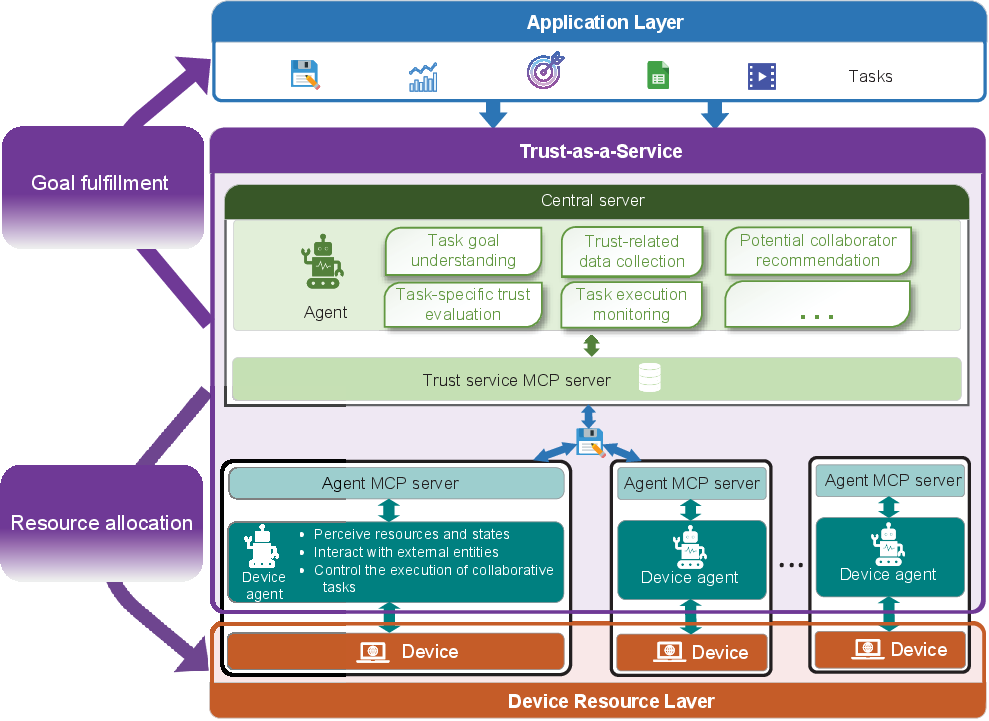}
\caption{The proposed TaaS serves as a bridge between the application layer and the device resource layer. Driven by task requirements from the application layer, the TaaS layer leverages trust as a guiding criterion to govern resource allocation at the device layer, ultimately ensuring the fulfillment of task goals.}
\label{framework}
\end{figure*}

\section{TaaS: Principles and Core Components}

% \textbf{Centralizing distributed data for comprehensive trust evaluation}:

\subsection{Principles}
TaaS encapsulates complex trust evaluation operations into a system-level, unified trust evaluation service available to all task owners, enabling autonomous, need-driven trust evaluation and trust-guided collaborator organization for effective task completion. Its core principles can be summarized in three aspects. First, by centralizing distributed semantic trust information on a reliable third-party server, the system can integrate information from diverse sources and provide a more comprehensive trust evaluation service to task owners, while avoiding the need for each task owner to directly query every potential collaborator. Second, TaaS interprets task-specific needs and translates them into relevant trust evaluation dimensions, enabling differentiated trust evaluation services for different tasks. This prevents tasks with different execution requirements from being evaluated using the same fixed criteria. Third, based on the need-dimension trust assessment results, TaaS dynamically organizes suitable collaborators into task-specific collaborative units for effective task completion.

\subsection{Core Components}
From a system architecture perspective, the proposed TaaS bridges the application layer and the device resource layer. Driven by task needs from the application layer, TaaS uses trust evaluation as the decision mechanism for coordinating suitable device resources at the device layer. As shown in Fig.~\ref{framework}, the proposed TaaS is built upon a central trust server, agentic AI deployed on both the server and individual devices, and MCP-based interfaces.

\textit{1) Agentic AI}: Agentic AI refers to AI systems designed to operate as autonomous agents that can pursue goals and take actions in dynamic environments. Each agent is capable of perceiving its environment, understanding task objectives, making decisions, and executing actions, enabling it to autonomously accomplish tasks in complex environments or collaborate with other agents to complete more complex tasks~\cite{11395598}.

In the proposed TaaS, the server-side agent serves as the core engine of the trust service. Upon receiving a task request, it performs a sequence of autonomous operations: (1) interpreting the task need description to derive need-specific trust evaluation dimensions, (2) identifying candidate devices according to task type and exposed capabilities, (3) performing trust evaluation along the selected need dimensions, and (4) returning semantic trust assessment results and candidate MCP access information for collaborator organization. On the device side, agents are deployed on participating IoT devices to monitor local states and capabilities, expose these capabilities to the system, and engage in collaborative task execution once selected. By deploying intelligent agents at both the server and device levels, the framework enables coordinated trust evaluation and efficient orchestration of collaborative execution across distributed devices.

\textit{2) MCP}: MCP is an open protocol introduced by Anthropic that standardizes how AI applications connect with external data sources and tools~\cite{10.1145/3796519}. In MCP, any entity that aims to expose its functionalities deploys an MCP server, which advertises its capabilities through three core primitives---tools (executable functions that perform actions), resources (data sources that provide contextual information), and prompts (reusable interaction templates). An MCP client can connect to an MCP server and dynamically discover these primitives through standardized interfaces (e.g., tools/list, resources/list), and subsequently invoke them at runtime without requiring predefined integrations~\cite{11406667}. This design decouples capability providers from capability consumers: the servers only need to describe what they offer through MCP primitives, while the clients can discover and utilize these capabilities on demand. Although MCP was originally designed for connecting LLM applications with external tools and data sources, its standardized capability exposure and dynamic discovery mechanisms are inherently extensible to broader collaborative scenarios where heterogeneous entities need to expose, discover, and utilize each other's capabilities.

In the proposed TaaS, MCP serves as the capability discovery and interaction layer connecting distributed agents. Device-side agents expose their capabilities through MCP interfaces, allowing the central server to dynamically discover available device resources and callable tools. Based on these discovered capabilities and the task needs interpreted by the server-side agent, the central server-side agent can perform need-driven trust evaluation. Meanwhile, the trust evaluation service provided by the server-side agent is also exposed through MCP interfaces, enabling task owners to access the service through a unified and standardized entry point. Through this design, MCP supports capability discovery, service invocation, and runtime interaction among distributed agents, thereby enabling scalable task-specific collaboration.

\section{Implementing TaaS via MCP-Aided Agentic AI}

This section presents the concrete implementation of TaaS through an agentic AI-driven framework built upon MCP. We first describe how the capabilities of both server-side and device-side agents are exposed externally via MCP servers. Using an example, we then illustrate how TaaS realizes need-driven trust evaluation service provisioning and trust-guided collaborator organization for effective task completion.

\begin{figure*}[t!]
\centering
\includegraphics[scale=1]{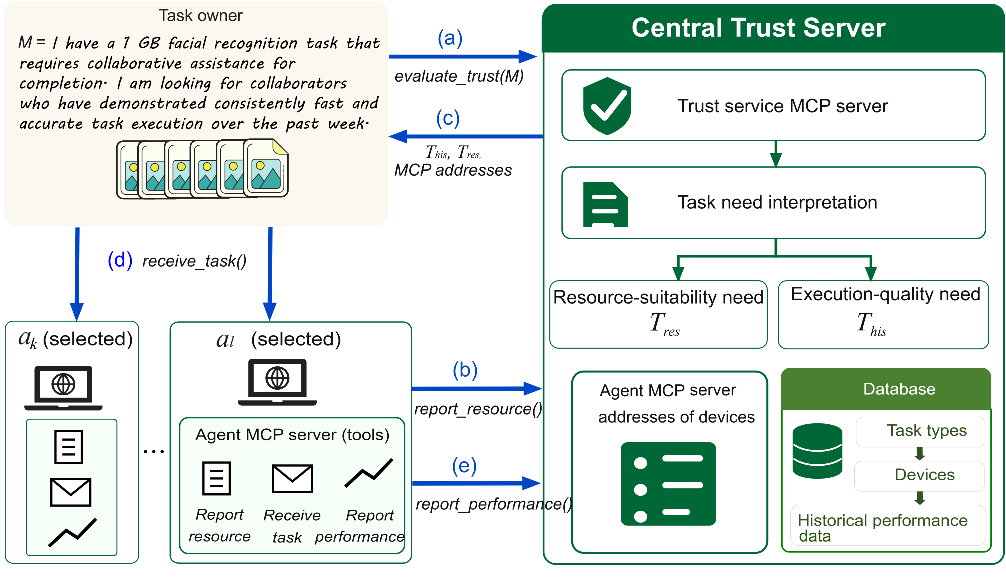}
\caption{The full implementation procedure of TaaS. (a) The task owner invokes the trust service by calling \texttt{evaluate\_trust()}; (b) the central server collects task-specific, semantic trust information from potential collaborators by invoking \texttt{report\_resource()}; (c) the central server returns the semantic trust results $T_{\text{his}}$ and $T_{\text{res}}$ of potential collaborators, along with the addresses of their agent MCP servers, to the task owner; (d) the task owner assigns the task to trusted collaborators by calling \texttt{receive\_task()}; (e) the central server monitors and collects performance data from collaborators during task execution by invoking \texttt{report\_performance()}.}
\label{workflow}
\end{figure*}

\subsection{System Initialization}

\textbf{Exposing server-side trust service}: The central server-side agent orchestrates the collaboration process tailored to each task and manages interactions with external entities.
In this implementation, the trust evaluation capability is externally exposed by encapsulating the server-side agent through the MCP protocol, forming the trust service MCP server. Within this server, the trust evaluation functionality is implemented as a remotely callable tool, \texttt{evaluate\_trust(\text{task description})}, which takes the task description as input and outputs need-dimension trust assessment results together with candidate collaborator information. Upon instantiation, the MCP server automatically generates a manifest describing this tool, providing a standardized description of the server-side agent's capability. The internal implementation of the tool is handled entirely by the agent, and other agents interacting with it do not require access to these details.

\textbf{Exposing device-side capability}:
Each agent deployed on a device manages internal device operations and external interactions. We encapsulate each device agent via the MCP protocol to expose its capabilities externally, forming an agent MCP server. In practice, devices have different hardware, software, and available resources, which leads to differences in the capabilities exposed by their device agents. For simplicity, we propose that all device agents provide three basic capabilities: reporting resource information, receiving tasks, and reporting task execution performance. These capabilities are exposed as three tools: \texttt{report\_resource()}, \texttt{receive\_task()}, and \texttt{report\_performance()}, as shown in Fig.~\ref{workflow}. It is worth noting that a device can dynamically extend its exposed capabilities through its agent MCP server. When a task requires richer device-side information or task-specific operations, the corresponding callable tools can be registered in the device-side agent MCP server and discovered through its manifest at runtime.

During system initialization, each device submits the address of its agent's MCP server to the central server agent and, in exchange, receives the address of the trust service MCP server. As a result, the central server agent is aware of all device agents' MCP server addresses, while every device agent is configured with the address of the trust service MCP server. This design enables MCP servers to be instantiated on demand rather than remaining continuously active. When a device joins the network, the types of tasks it supports are registered with the central server, providing the basis for subsequent candidate discovery in need-driven trust evaluation. This registration does not establish a binding relationship between the device and the central server.

\subsection{Task Owner-Side Task Generation}

The system consists of a set of devices $A=\{a_1,\dots,a_N\}$. Each device can generate computation tasks of different types and request the trust server to select trusted collaborators for task execution. Suppose that device $a_i$, acting as the task owner, generates a facial recognition task. The task is characterized by its data size and the task owner's collaborator selection objectives. The agent of device $a_i$ first converts the task information and the task owner's objectives into a textual description $M$, such as: ``I have a 1 GB facial recognition task that requires collaborative assistance for completion. I am looking for collaborators that can process the task responsively and have demonstrated consistently fast and accurate task execution over the past week.''
Subsequently, the agent of device $a_i$ submits this task need description $M$ to the server-side agent via the MCP protocol. Specifically, device $a_i$'s agent first establishes a connection to the trust service MCP server using the preconfigured address. Upon establishing the connection, the agent of device $a_i$ retrieves the manifest describing the tools provided by the trust service MCP server. After verifying the availability of the trust evaluation tool \texttt{evaluate\_trust(\text{task description})}, the agent of device $a_i$ invokes the tool, passing the task need description $M$ as an argument.

\subsection{Need-Driven Trust Evaluation Service Provisioning on the Server}

% Upon receiving a task description, the central server agent adaptively determines and autonomously executes a task-specific trust evaluation process. 

After receiving the task description from the task owner, the trust server provisions the trust evaluation service according to the task-specific requirements. Based on the tailored evaluation process, the server-side agent determines the relevant evaluation criteria and identifies the devices that can potentially serve as collaborators for the task.

\textbf{Task need interpretation}: To realize need-driven service provisioning, the central server agent interprets the task description $M$ to identify the task needs that are relevant to trust evaluation. Since different tasks require different collaborator properties, these task needs must be translated from high-level natural-language objectives into measurable evaluation indicators and scopes. In this example, the task is recognized as a facial recognition task with a data size of 1 GB. The need for ``consistently fast and accurate task execution over the past week'' corresponds to the execution-quality need, represented by indicators such as task processing speed and task completion accuracy, with the past week serving as the evaluation scope. Meanwhile, the need to process a 1 GB task with sufficient responsiveness corresponds to the resource-suitability need, represented by indicators such as available storage capacity and CPU capability. In this way, the extracted task needs define a task-specific trust evaluation space with named need dimensions, and the server can subsequently evaluate potential collaborators along these dimensions according to the actual requirements of the task.

\vspace{-0.03 in}
\textbf{Trust evaluation along the execution-quality need dimension}: Based on the task need interpretation, the execution-quality need of the facial recognition task is represented by the task processing speed and task completion accuracy evaluated over the past week. This need dimension determines not only what evidence should be used for trust evaluation, but also which devices should be considered. Specifically, the server-side agent first uses the task type to query the local task-support registry and obtain devices that support facial recognition tasks, thereby forming the candidate set relevant to the current task. It then retrieves the historical execution records of these candidates from the local database, which is organized by task type, device identity, and historical performance data, as shown in Fig.~\ref{workflow}. Guided by the execution-quality need, the server-side agent extracts only the evidence associated with facial recognition tasks, the past-week time scope, task processing speed, and task completion accuracy. The generated semantic trust assessment therefore reflects whether each candidate can satisfy this need dimension, instead of providing a general historical trust score. In this example, devices $a_j$, $a_k$, and $a_l$ satisfy the execution-quality need, and their assessments are represented as $T_{\text{his}}=$\{\{``device'': ``$a_j$'', ``facial recognition-specific execution quality'': ``task processing speed is 10 MB/second, task completion accuracy is 100\%''\}, \{``device'': ``$a_k$'', ``facial recognition-specific execution quality'': ``task processing speed is 12 MB/second, task completion accuracy is 100\%''\}, \{``device'': ``$a_l$'', ``facial recognition-specific execution quality'': ``task processing speed is 12 MB/second, task completion accuracy is 100\%''\}\}.

\vspace{-0.03 in}
\textbf{Trust evaluation along the resource-suitability need dimension}: The resource-suitability need extracted from the task description specifies whether a candidate device has sufficient resources to support the current task. For the facial recognition task, this need is represented by available storage capacity and CPU capability, which are derived from the 1 GB data size and the requirement for responsive processing. Given the candidate devices, the server-side agent performs trust evaluation along this need dimension by collecting only the resource information required by the task, rather than querying all resource states of each device. Specifically, using the preconfigured agent MCP server addresses of devices $a_j$, $a_k$, and $a_l$, the server-side agent establishes connections to these devices and retrieves their available tools. Guided by the resource-suitability need, it invokes the \texttt{report\_resource()} tool with the natural-language request ``Please report your currently available storage capacity and CPU capability for supporting a 1 GB facial recognition task that requires responsive processing.'' Each device agent then collects the requested resource states and reports them back to the server-side agent. Based on the returned resource information, the server-side agent generates a semantic trust assessment indicating whether each candidate satisfies the resource-suitability need. The evaluation results are represented as $T_{\text{res}}=$ \{\{``device'': ``$a_j$'', ``facial recognition-specific resource suitability'': ``CPU is 2 GHz (moderate processing speed), and the available storage is 4 GB ($>$ 1 GB required)''\}, \{``device'': ``$a_k$'', ``facial recognition-specific resource suitability'': ``CPU is 6 GHz (high processing speed), and the available storage is 8 GB ($>$ 1 GB required).''\}, \{``device'': ``$a_l$'', ``facial recognition-specific resource suitability'': ``CPU is 6 GHz (high processing speed), and the available storage is 4 GB ($>$ 1 GB required)''\}\}. After completing the trust assessments along the extracted need dimensions, the central server agent returns $T_{\text{his}}$, $T_{\text{res}}$, and the agent MCP server addresses of devices $a_j$, $a_k$, and $a_l$ to the task owner as the output of the \texttt{evaluate\_trust()} tool.

\subsection{Trust-Guided Collaborator Organization for Task Completion}

\textbf{Trusted collaborator selection}: After receiving the trust evaluation results from the trust server, the task owner $a_i$ uses them as the decision basis for forming a task-specific collaborative unit. Specifically, $T_{\text{his}}$ indicates the degree to which each candidate satisfies the execution-quality need, while $T_{\text{res}}$ indicates the degree to which it satisfies the resource-suitability need. By jointly considering these need-dimension trust assessments, the task owner selects the devices that best match the actual requirements of the facial recognition task. In this example, devices $a_j$, $a_k$, and $a_l$ all satisfy the execution-quality need and provide sufficient storage capacity. However, device $a_j$ only has moderate CPU capability, making it less suitable for the resource-suitability need of this task. Therefore, $a_i$ selects devices $a_k$ and $a_l$ as collaborators.

\textbf{MCP-based collaborator engagement}: After the collaborators are selected, the task owner engages them through their agent MCP servers. Using the returned MCP server addresses, $a_i$ establishes connections to devices $a_k$ and $a_l$ and retrieves their available tools. After confirming that the \texttt{receive\_task()} tool is available, $a_i$ decomposes the 1 GB facial recognition task into two subtasks according to the selected collaborators' capabilities and the execution requirements of the task. Each subtask is assigned to one collaborator through \texttt{receive\_task()}, together with the task type, input data partition, and expected output format. In this way, the trust evaluation results are translated into concrete task allocation actions through standardized MCP tool invocation.

\textbf{Execution monitoring and trust feedback}: During task execution, the central server agent continues to support the collaboration lifecycle by monitoring the collaborators' runtime performance. Specifically, it invokes the \texttt{report\_performance()} tool exposed by the collaborator agents to obtain execution feedback, such as task progress, processing speed, completion status, and runtime resource conditions. If a collaborator shows significant performance degradation or can no longer satisfy the task needs, the central server agent notifies the task owner and coordinates collaborator replacement or task reassignment. After the collaborators complete their assigned subtasks, the task owner aggregates the returned results to obtain the final facial recognition output. Meanwhile, the execution feedback is recorded by the central server as historical performance data for future need-driven trust evaluation. Finally, the MCP connections associated with the trust server and the selected collaborators are closed, and the occupied resources are released.

\section{Results Analysis}

To validate the proposed approach, we construct a collaborative system using multiple devices, including MacBook, Dell Latitude 5280, and Dell EMC 5200. All agents are implemented based on the OpenAI Agents SDK, while MCP servers are developed using FastMCP.

\begin{figure}[!]
\centering
\includegraphics[scale=0.8]{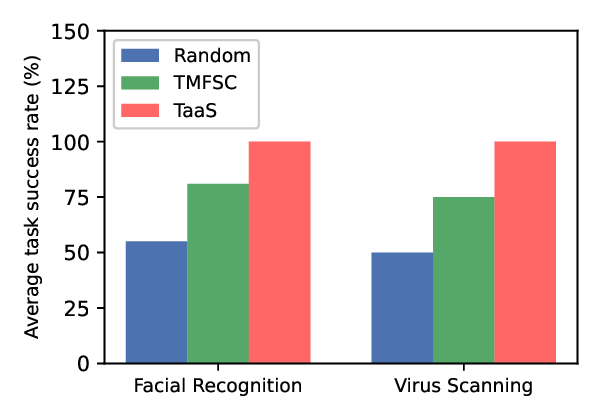}
\caption{The proposed TaaS achieves consistently higher task success rates across different tasks by enabling accurate, task-specific collaborator selection.}
\label{task_success}
\end{figure}

\begin{figure}[!]
\centering
\includegraphics[scale=0.8]{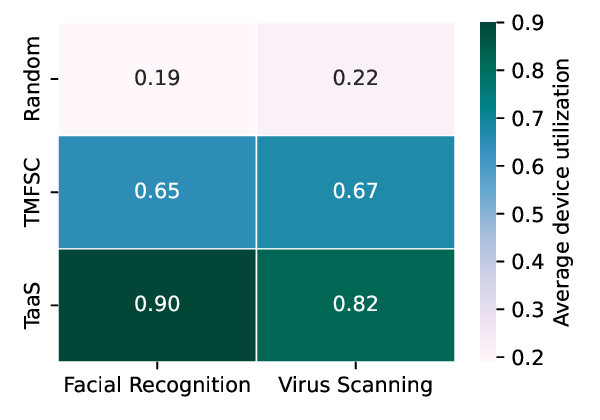}
\caption{The proposed TaaS achieves the highest device utilization across tasks, demonstrating that task-specific collaboration enables more efficient and effective resource usage.}
\label{heatmap}
\end{figure}

Fig.~\ref{task_success} presents a comparison of the average task success rates for random, TMFSC~\cite{10944812}, and the proposed method across two task types: facial recognition and virus scanning~\cite{11296817}. It can be observed that the random method exhibits the lowest success rates, indicating its limited reliability due to random selection. TMFSC achieves moderate performance, while our proposed TaaS reaches a 100\% success rate for both tasks. This demonstrates that by accurately understanding task objectives and leveraging task-specific historical performance and resource evaluations, the proposed TaaS can effectively select the most suitable devices. The result is consistent with the design of TaaS, where trust evaluation is performed according to the needs of the current task before collaborator selection.

Fig.~\ref{heatmap} illustrates the average device utilization across different algorithms and task types, where device utilization is defined as the ratio of the number of devices that actually execute the task to the total number of devices involved in each task. It can be observed that the random method exhibits the lowest utilization, indicating significant resource waste, while TMFSC shows moderate improvement. However, the proposed TaaS achieves the highest device utilization across both tasks. These results demonstrate that the proposed TaaS enables more efficient utilization of devices, thereby reducing resource waste and facilitating efficient collaboration. This improvement comes from selecting collaborators whose capabilities and resources are aligned with the task needs, rather than involving poorly matched devices.

\begin{figure}[!t]
\centering
\includegraphics[scale=0.8]{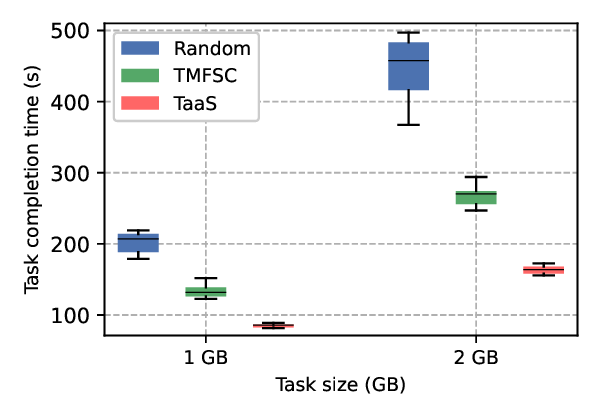}
\caption{The proposed TaaS achieves the shortest and most stable task completion times across different task sizes, outperforming the random and TMFSC methods in both efficiency and consistency.}
\label{task_time}
\end{figure}

Fig.~\ref{task_time} illustrates the distribution of task completion times for different algorithms under two task sizes--1 GB and 2 GB--for facial recognition tasks. The boxplots are used to illustrate the variability of each algorithm, while the median provides a robust measure of central tendency for each algorithm's completion time. It can be observed that the random method yields the longest completion time, with a wider box and longer whiskers, indicating significant performance variability and poor stability. The TMFSC method improves both task completion time and stability, although noticeable fluctuations still exist. In contrast, the proposed TaaS achieves the shortest task completion time under both task sizes, with the most compact boxplots and the smallest variation range. This indicates that the proposed TaaS can select collaborators according to the interpreted resource-suitability and execution-quality needs, thereby achieving superior execution stability.

Overall, the proposed TaaS outperforms the baseline approaches in terms of both execution efficiency and stability, validating the effectiveness of the task-specific collaborative mechanism in optimizing task scheduling and improving system performance. The consistent gains across success rate, device utilization, and completion time further show the benefit of aligning trust evaluation with task needs.

\section{Future Directions}
While the proposed TaaS framework establishes a solid foundation for trust-driven collaborative systems, several promising directions remain open for future exploration.

\textbf{Decentralized and federated TaaS architectures}: Extending TaaS beyond centralized deployment, future work will explore distributed and federated trust management architectures in which trust services are collaboratively maintained across multiple nodes, reducing dependence on a single trust server and improving resilience and scalability in large-scale systems.

\textbf{Privacy-preserving trust evaluation}: Future work will investigate the integration of privacy-preserving mechanisms--such as federated learning and differential privacy--into the proposed TaaS, enabling trustworthy and accurate collaborator assessment while limiting the exposure of sensitive device-level data.

\textbf{Embedding-enhanced trust environment construction}: Future work will also consider incorporating embedding techniques into the construction of the trust environment, particularly for data transmission scenarios. By representing heterogeneous transmission contexts, interaction records, and device states in a semantic embedding space, the trust server may capture richer contextual relationships among trust-related factors, thereby improving the semantic representation and accuracy of need-driven trust evaluation within the proposed TaaS framework.

\section{Conclusion}

This paper has presented TaaS, a novel paradigm that elevates trust evaluation from a fragmented, device-level function to a unified, system-wide service, enabling task-specific and trustworthy operation across collaborative IoT systems. We implemented TaaS through an MCP-aided agentic AI framework. The framework provisions trust evaluation services to task owners by aligning diverse task requirements with available system capabilities, enabling dynamic collaborator orchestration for effective task completion. By interpreting task needs, evaluating candidates along relevant trust dimensions, and engaging selected devices through MCP servers, TaaS links trust evaluation with concrete task execution. Evaluation results demonstrate that the framework supports accurate collaborator selection, resilient task execution, and efficient resource utilization under diverse scenarios. These results confirm that need-driven trust evaluation can improve collaborator selection accuracy, device utilization, and completion-time stability compared with the baseline methods.

% Generated by IEEEtran.bst, version: 1.14 (2015/08/26)

% \vspace{0.04 in}
% \footnotesize
% \bibliographystyle{IEEEtran}
% \bibliography{IEEEabrv, bibliography_file.bib} 
% \normalsize

\end{document}